\documentclass{PoS}

\title{Valence transversities: the collinear extraction.}

\ShortTitle{Valence transversities: the collinear extraction.}

\author{\speaker{A. Courtoy}%
         \thanks{Also associated to INFN-LNF, Frascati}\\
        IFPA, AGO Department, University of Li\`ege, Belgium\\
        E-mail: \email{aurore.courtoy@ulg.ac.be}}
        
\author{ Alessandro Bacchetta\\
        University of Pavia, Italy \& INFN-Sezione di Pavia\\
        E-mail: \email{alessandro.bacchetta@unipv.it}}
        
\author{ Marco Radici\\
        INFN-Sezione di Pavia, Italy\\
        E-mail: \email{marco.radici@pv.infn.it}
        }

\abstract{In these proceedings, we propose an extraction of the valence transversity parton distributions. Based on an analysis of pion-pair production in deep-inelastic scattering off transversely polarized targets, this extraction of transversity is performed in the framework of collinear factorization. The recently released data for proton and deuteron targets at HERMES and COMPASS allow for a flavor separation of the valence transversities, for which we give a complete statistical study.}

\FullConference{XXI International Workshop on Deep-Inelastic Scattering and Related Subjects\\
		 22-26 April, 2013\\
		 Marseilles, France}

\begin{document}

\section{Introduction}

The transversity parton distribution function (PDF) is poorly known with respect to the 2 other leading-twist distributions. In consequence of its chiral-odd nature, transversity  is not observable from fully inclusive Deep Inelastic Scattering (DIS).  It can be measured in processes that either have two hadrons in the initial state, {\it e.g.}, proton-proton collision ; or one hadron in the initial state and at least one hadron in the final state, {\it e.g.}, semi-inclusive DIS (SIDIS).

It was considering the latter kind of processes --more precisely, single-hadron SIDIS-- that  the valence transversities were extracted for the very first time by the Torino group~\cite{Anselmino:2008jk,Anselmino:2013vqa}. The main shortcoming in analyzing such processes, and the deriving outcomes, lies in the factorization framework they must obey. In effect, most single-hadron SIDIS processes involve Transverse Momentum Dependent functions (TMDs), the QCD evolution of which  is still under active debate in the community. The impact of  evolution on the transversity TMD has been recently studied~\cite{Bacchetta:2013pqa} in the so-called CSS framework.

An alternative to the evolution related issue arose thanks to the so-called Dihadron Fragmentation Functions (DiFFs)~\cite{Jaffe:1997hf,Radici:2001na}, that are defined in a collinear factorization framework. Here, we will give an overview of the technical strategy used to extract the valence collinear transversities.

In this contribution to the proceedings, we give  a parameterization of the valence transversities together with the error analyses~\cite{Bacchetta:2012ty}. The fitting procedure is based on the knowledge of DiFFs from a previous analysis~\cite{Courtoy:2012ry}. The outcome confirms our first extraction of the proton transversity~\cite{Bacchetta:2011ip}. Results and analysis of first principle properties such as the tensor charge are discussed. 


\section{Two-Hadron Semi-Inclusive DIS}

We consider two-hadron production in DIS, {\it i.e.},  the process
\begin{equation}
\ell(l) + N(P) \to \ell(l') + H_1(P_1)+H_2(P_2) + X \enspace,
\end{equation}
where $\ell$ denotes the beam lepton, $N$ the nucleon target, $H_1$ and $H_2$
the produced hadrons,  
and where four-momenta are given in parentheses. For a transversely polarized target, the cross section is related to the simple product of the (collinear) transversity distribution function and the chiral-odd DiFF, denoted as
$H_1^{\sphericalangle\,q}$~\cite{Radici:2001na}. The latter describes the correlation between the 
transverse polarization of the fragmenting quark with flavor $q$, related to $\phi_S$, and the azimuthal 
orientation of the plane containing the momenta of the detected hadron pair, related to $\phi_R$. In such a process, the ``transverse" reference comes from the  projection of the relative momentum of the hadron pair 
onto the plane tranverse  to the jet axis. This leads to an asymmetry in $\sin(\phi_S+\phi_R)$.

Hence, in such a collinear framework, we can single out the DiFF contribution to the SIDIS asymmetry from the $x$-dependence coming from the transversity PDF. 
In particular, the $x$ behavior of $h_1(x)$ is simply given by integrating the numerator of the asymmetry over the $(z, M_h)$-dependence. This procedure then defines the quantities $n_q $ and $n_q^\uparrow $ that are the integrals of, respectively, the unpolarized and chiral-odd DiFFs.
The relevant single-spin asymmetry in SIDIS here can be  expressed in terms of the integrated DiFFs. It reads
\begin{eqnarray}
A_{\mbox{\tiny SIDIS}} (x, Q^2) &=&-C_y\,\frac{\sum_q\, e_q^2\, h_1^q(x, Q^2) n_q^{\uparrow}(Q^2)}{\sum_q\, e_q^2\, f_1^q(x, Q^2) n_q(Q^2)}\enspace,
  \label{eq:asy_dis}
\end{eqnarray}
where the sum runs over all the quark flavors.  The depolarization factor $C_y=1$ for COMPASS data and  $C_y\approx
\frac{(1-\langle y \rangle)}
       {(1-\langle y \rangle +\langle y \rangle^2/2)}$
for HERMES data. 
       
Given  the asymmetry data and using $f_1(x)$ from available sets, the remaining unknows in eq.~(\ref{eq:asy_dis}) are the transversities and the integral of the DiFFs. The latter are extensively described in Refs.~\cite{Bacchetta:2012ty,Courtoy:2012ry}. We here only need to recall that we used isospin symmetry and charge conjugation to relate the polarized or unpolarized DiFFs of different flavors~\cite{Bacchetta:2006un}, what leaves only $n_u, n_d, n_s$ and $n_u^\uparrow$ to the purpose.


\section{The extraction from a collinear framework}
\label{sec:tra_colli}
 
 Starting with the expression~(\ref{eq:asy_dis}) for the asymmetry, we discuss  the  extraction of the valence transversities from combining data on both proton and deuteron target from COMPASS and proton target from HERMES.
The main theoretical constraint we have is Soffer's 
inequality~\cite{Soffer:1994ww},  
\begin{equation} 
2|h_1^q(x; Q^2)| \leq | f_1^q(x; Q^2) + g_1^q(x; Q^2)| \equiv 2\,\mbox{\small SB}^q(x; Q^2) \; .
\label{e:soffer}
\end{equation} 
If the Soffer bound is fulfilled at some 
initial $Q_0^2$, it will hold also at higher 
$Q^2 \geq Q_0^2$.
We impose this positivity bound by multiplying the functional form by the corresponding Soffer
bound at the starting scale of the parameterization, $Q_0^2=1$ GeV$^2$. 
The implementation of the 
Soffer bound depends on the choice of the unpolarized and helicity PDFs. We use  
the MSTW08 set~\cite{Martin:2009iq} for the unpolarized PDF, combined to the 
DSSV parameterization~\cite{deFlorian:2009vb} for the helicity distribution, at the 
scale of $Q_0^2=1$ GeV$^2$. Our analysis was carried out at LO in $\alpha_S$.
The functional form that ensues 
for the valence transversity distributions reads,
\begin{equation} 
x\, h_1^{q_V}(x; Q_0^2)=
\tanh \left[ x^{1/2} \, \bigl( A_q+B_q\, x+ C_q\, x^2+D_q\, x^3\bigr)\right]\, 
\left[ x\, \mbox{\small SB}^q(x; Q_0^2)+x\, \mbox{\small SB}^{\bar q}(x; Q_0^2)\right]  .
\label{e:funct_form}
\end{equation} 
The hyperbolic tangent is such that the Soffer bound is always fulfilled. 
The functional form is very flexible and can contain up to three nodes but the low-$x$ behavior is  determined by the $x^{1/2}$ term, which is imposed 
by hand in order to insure the integrability of the transveristy PDF. We have considered different sets of  parameters. However, in these proceedings, we will give results for the set which contains 6 parameters, i.e.,  $D_u=D_d=0$. We call it the {\it flexible scenario}.  

The fit, and in particular the error analysis, 
was carried out in two different ways: using the
standard Hessian method and using a Monte Carlo approach. The latter is inspired from the work of
the NNPDF collaboration, {\it e.g.}, \cite{Forte:2002fg}, 
although our results are not based on a neural-network fit. The approach consists in 
creating $N$ replicas of the data points, shifted by a Gaussian noise with the same variance as the 
measurement. Each replica, therefore, represents a possible outcome of an 
independent experimental measurement. Then, the standard minimization procedure is applied to each replica separately as explained in ref.~\cite{Bacchetta:2012ty}.

\begin{figure}
\begin{center}
\includegraphics[width=0.49\textwidth]{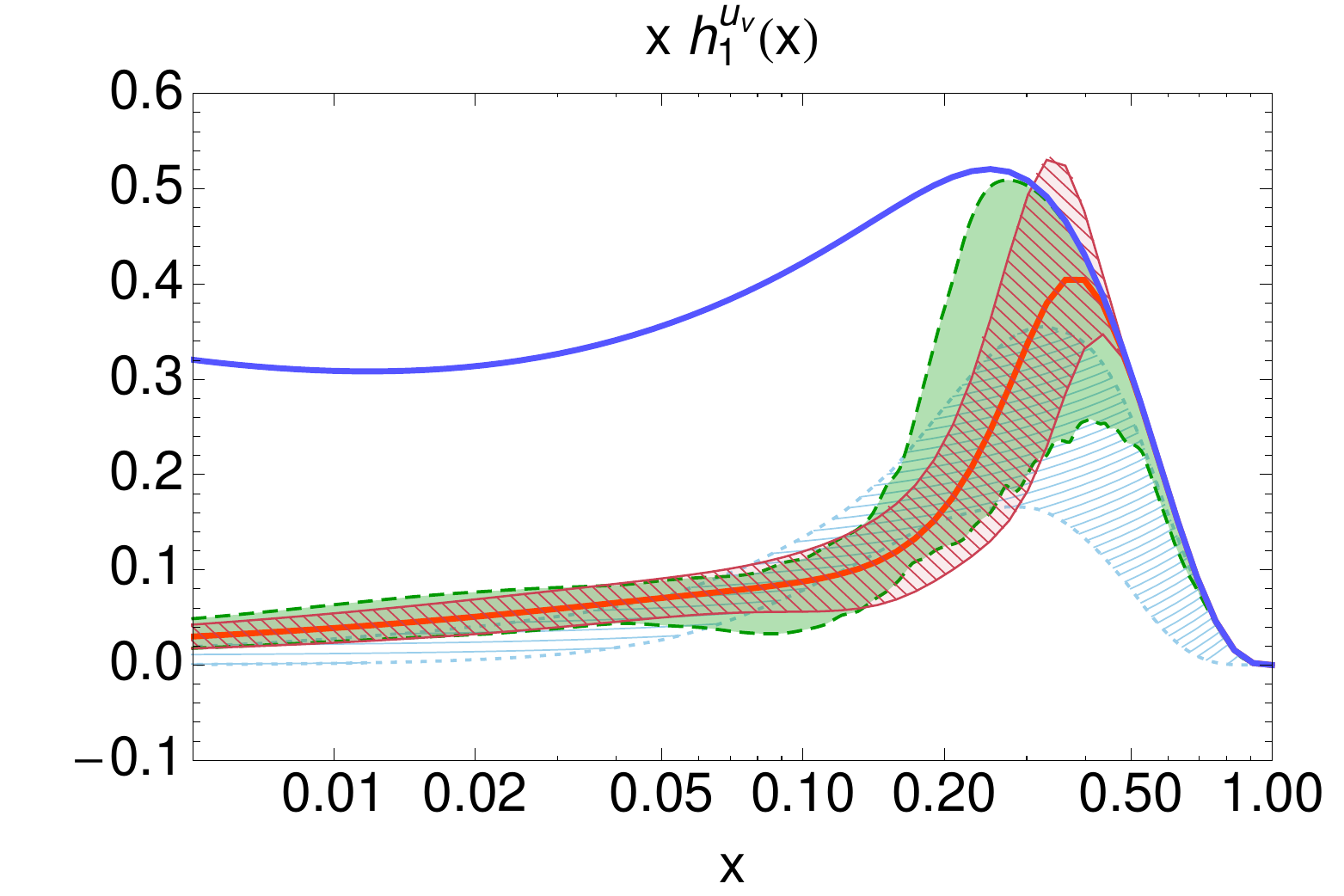}
\includegraphics[width=0.49\textwidth]{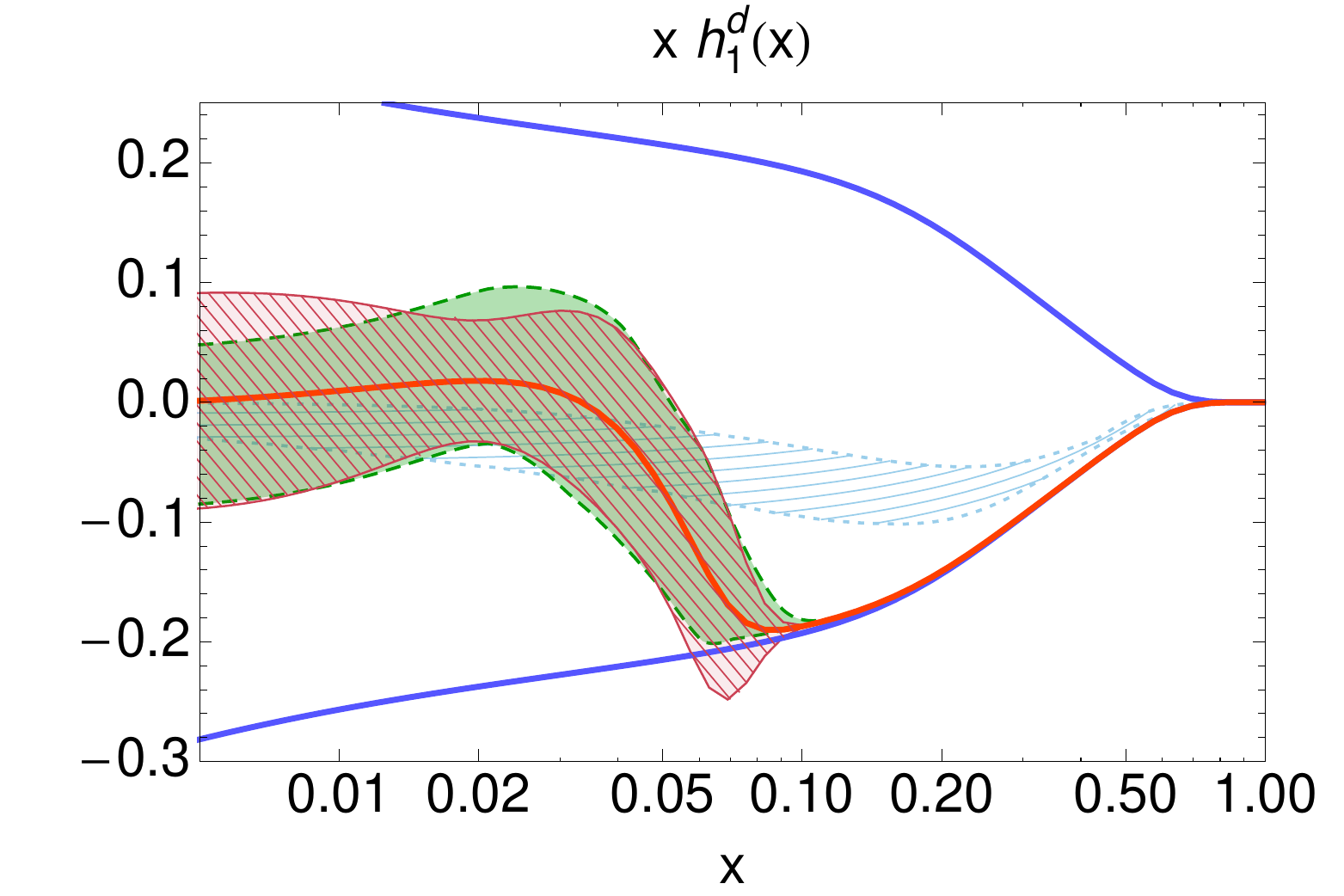}
\caption{\label{fig:flex_val} 
The valence transversities at $Q^2=2.4$ GeV$^2$. The thick solid  blue lines are the Soffer bound. The uncertainty band (red stripes)  with solid 
    boundaries is the best fit in the standard approach at $1\sigma$, whose central 
    value is given by the central thick solid red line. The uncertainty band (plain green) with dashed 
    boundaries is the $68\%$ of all fitting replicas obtained in the Monte Carlo approach. 
    As a comparison, the uncertainty band (blue stripes) with short-dashed boundaries is the transversity 
    extraction from the Collins effect~\cite{Anselmino:2008jk}.}
\end{center}
\end{figure}

We show the resulting functional form for the valence transversities at $Q^2=2.4$ GeV$^2$ of both fitting procedures for the {\it flexible scenarios} on fig.~\ref{fig:flex_val}. In the standard Hessian method, the $\chi^2/$d.o.f. is  $1.12$ for the {\it flexible} scenario, while the average $\chi^2$
is  $\chi^2/$d.o.f.$= 1.56$ for  100 replicas.

The uncertainty bands 
in the standard and Monte Carlo approaches are quite similar. The main difference 
is that in the former case the boundaries of the band can occasionally cross the Soffer 
bound. 
This is due to the fact that the hypothesis under which we can use the standard Hessian approach are no longer valid in this limit.
On the contrary, in the Monte Carlo approach each replica is built such that it 
never violates the Soffer bound; the resulting $68\%$ band is always within 
those limits. We observe that the standard approach for the $u_v$ tends
to saturate the Soffer bound at $x\sim 0.4$. In the Monte Carlo approach, some of the replicas saturate the bound already 
at lower values. This happens in both approaches at values of $x$ for which no data are available. 
In the  range  where data exist, 
our results are compatible with the Torino parametrization of 
transversity~\cite{Anselmino:2008jk,Anselmino:2013vqa}. The only 
source of discrepancy lies in the range $0.1< x <0.16$ for the
valence down quark\footnote{ It might be due to  two experimental data for the deuteron target.}. 

A qualitative comparison between our results  and the 
available model predictions shows that the extracted transversity 
for the up quark is smaller than most of the model calculations at intermediate 
$x$, while it is larger at lower $x$. The down 
transversity is much larger in absolute value than all model calculations at 
intermediate $x$ (as observed before, this is due to the deuterium data points), 
while the error band is too large to draw any conclusion at lower $x$.
\\

One direct appliation of the extraction of transversity is the evaluation of the tensor charge, a fundamental quantity 
of hadrons at the same level as the vector, axial, and scalar charges. 
The tensor charge remains at the moment largely unconstrained. 
There is no sum rule related to the tensor current, due to the property of the
anomalous dimensions governing the QCD evolution of transversity. 
The contribution of a flavor $q$ to the tensor charge is defined as,
\begin{equation}
\delta q(Q^2)=\int \, dx\, h_1^{q_v}(x; Q^2) \; .
\label{e:tensor}
\end{equation}
The region of validity of our fit is restricted to the experimental data
range. We can therefore give a reliable estimate for the tensor charge 
truncated to the interval $x \in [0.0064,0.28]$. For the {\it standard flexible} functional form, we find
\begin{equation}
\delta u_{tr}(Q^2)=0.29 \pm 0.13\quad,\quad  \delta d_{tr}(Q^2)=-0.26 \pm 0.22\quad,
\end{equation}
which is compatible with what we obtain with the other functional forms. However, once we try to integrate over the all range $0<x<1$, the results do not agree as much. It is yet another sign that we are lacking information at low and large $x$ values.

\section{Conclusions}

We have proposed a parametrization of the valence transversities in a collinear framework. It relies on the knowledge of  DiFFs, which we have studied and parametrized in previous works. The transversity analysis is driven in two different statistical approaches: a standard and a Monte Carlo-like one.

We can conclude that,
outside the kinematical range of experiments, the lack of data reflects itself
in a large uncertainty in the parametrization, reflecting in the  tensor charge integrated over theoretical range. 
This illustrates  the need for new large-$x$ data in order to 
reduce the degree of uncertainty in the knowledge of transversity. Also, since the tensor charge is defined as the integral of the transversity PDF on its support, the shape of the valence transversities, at both low and large-$x$, plays a key role in determining the tensor charge. Data from CLAS12 and SoLID at Jefferson Lab (expected in the next years) should help unfolding the situation.

\end{document}